\documentclass[twoside,preprintnumbers,amsmath,amssymb,pacs,shownopacs,nofootinbib]{revtex4}
\usepackage{graphicx,subfigure}
\usepackage{braket,euscript,enumitem}
\usepackage{fancyvrb}
\usepackage{physics,slashed}
\usepackage{xcolor}

\renewcommand{\d}{\ensuremath{\mathrm{d}}}

\newcommand{\p}{\partial}

\allowdisplaybreaks[2]
\begin{document}

\title{{\bf A fresh look at the (non-)Abelian Landau-Khalatnikov-Fradkin transformations}}

\author{T.~De Meerleer$^{a}$, D.~Dudal$^{a,b}$, S.~P.~Sorella$^c$}
\email{tim.demeerleer@kuleuven.be, david.dudal@kuleuven.be, silvio.sorella@gmail.com}
\affiliation{${}^a$ KU Leuven Campus Kortrijk -- KULAK, Department of Physics, Etienne Sabbelaan 53 bus 7657, 8500 Kortrijk, Belgium\\ ${}^b$ Ghent University, Department of Physics and Astronomy, Krijgslaan 281-S9, 9000 Gent, Belgium \\${}^c$ Instituto de F\'isica Te\'orica, Rua S\~ao Francisco Xavier 524, 20550-013, Maracan\~a, Rio de Janeiro, Brasil}

\author{P.~Dall'Olio$^d$, A.~Bashir$^d$}
\email{odiotin@hotmail.com, adnan@ifm.umich.mx}
\affiliation{${}^d$ Instituto de F\'{i}sica y Matem\'{a}ticas,
Universidad Michoacana de San Nicol\'{a}s de Hidalgo, Edificio
C-3, Ciudad Universitaria, Morelia, Michoac\'{a}n 58040,
M\'{e}xico}

\begin{abstract}
The Landau-Khalatnikov-Fradkin transformations (LKFTs) allow to
interpolate $n$-point functions between different gauges. We first
offer an alternative derivation of these LKFTs for the gauge and
fermions field in the Abelian (QED) case when working in the class
of linear covariant gauges. Our derivation is based on the
introduction of a gauge invariant transversal gauge field, which
allows a natural generalization to the non-Abelian (QCD) case of
the LKFTs. To our knowledge, within this rigorous
formalism, this is the first construction of the LKFTs beyond
QED. The renormalizability of our setup is guaranteed to all
orders. We also offer a direct path integral derivation in the
non-Abelian case, finding full consistency.
\end{abstract}
\maketitle

\section{Introduction}

When we study strong color interaction, namely quantum
chromodynamics (QCD), we start from the most basic fields, namely the quarks, gluons and also the Faddeev-Popov ghosts in covariant gauges. Due to the infrared enhancement  of the strong coupling constant, perturbation theory
alone is unable to provide a description of the observable
hadronic world made up of quarks and gluons.
Therefore, the need for non-perturbative approaches
arises, requiring a radically different
treatment of these interactions.

In the continuum formulation, gauge fixing is
required to warrant computations, whatever non-perturbative
scheme one has in mind. However, QCD remains a gauge theory,
meaning physically observable quantities should not
depend on what gauge is actually chosen to carry out the
computation. In this article, we concern ourselves with linear
covariant gauges, with the Landau gauge as a special case thereof.

Within the functional approach of Dyson-Schwinger equations (DSEs)
\cite{Roberts:1994dr,Alkofer:2000wg,Aguilar:2008xm,Binosi:2009qm,Fischer:2008uz,Boucaud:2008ky,Boucaud:2011ug,Maas:2011se,Bashir:2012fs,Eichmann:2016yit} or functional renormalization group equations \cite{Pawlowski:2005xe,Cyrol:2017ewj}, one is
confronted with an infinite tower of non-linear coupled equations
with an ever-increasing order of $n$-point correlation functions.
This is of course unamenable to computation, so a sacrifice must
be made: the tower is truncated and some of the necessary low
order $n$-point correlation functions are introduced via a
sensible Ans\"atze preserving some key features of a
gauge field theory. Much care is generally taken as regards the
low energy constraint of chiral symmetry, namely the
axial vector Ward identity, and the pattern in which this symmetry
is dynamically broken. The corresponding low energy
Goldberger-Treiman relations provide an intimate connection
between the quark propagator and the Bethe-Salpeter amplitudes of
the corresponding bound state. It is of paramount importance, not
only to get the correct QCD spectrum for low lying mesons, but
also to study corresponding elastic and transition form factors
which have come under immense experimental and theoretical
scrutiny in the last few
years~\cite{Huber:2008id,Aubert:2009mc,Uehara:2012ag,Chang:2013nia,Raya:2015gva,Raya:2017ggu}.

 However, the constraints of gauge covariance are not always fully
implemented. These constraints manifest themselves not only in
terms of Slavnov-Taylor identities but also as generalized
Landau-Khalatnikov-Fradkin transformations (LKFTs) which are less
studied, except for the Abelian (QED) case, see
e.g.~\cite{Roberts:1994dr,Burden:1993gy,Bashir:2002sp,Bashir:2009fv}
and recent works such as
\cite{Jia:2016udu,Jia:2016wyu,Bermudez:2017bpx}. In principle, if
gauge covariance is manifest, transforming the $n$-point functions
in one gauge to those in another gauge, will have no consequence
whatsoever on physical observables computed from these $n$-point
functions. However, as soon as model-building is done in some
particular gauge, there can be conflicts with gauge (BRST)
invariance which can lead to uncontrollable gauge parameter
dependence filtering into physical quantities. For
example, one can expect such a thing to happen if models specific
to Landau gauge were to be used in other gauges without appropriate gauge modifications.

In fact, most functional studies are restricted to the Landau
gauge case, because of its interesting properties, such as it
being a fixed point of the renormalization group and the fact that it is accessible also to gauge-fixed lattice QCD
studies
\cite{Cucchieri:2007md,Cucchieri:2007rg,Cucchieri:2008fc,Bogolubsky:2009dc,Bowman:2007du,Maas:2008ri,Silva:2004bv,Oliveira:2012eh,Dudal:2013yva,Furui:2003jr,vonSmekal:2009ae}.
However, during the course of past few years, we have witnessed an
increased activity in extending both functional and numerical
lattice efforts to general linear covariant gauges. In the long
run can, it will lead to a better understanding  of how to extract
truly gauge invariant physical information in a gauge fixed
context
\cite{Cucchieri:2009kk,Cucchieri:2011aa,Bicudo:2015rma,Huber:2015ria,Aguilar:2015nqa,Aguilar:2016ock,Siringo:2014lva,Siringo:2015gia,
Capri:2015ixa,Capri:2015nzw,Capri:2016aqq,AminAmin,Capri:2016gut,Capri:2017bfd}.
That such  a goal is far from  being trivial has been illustrated
even in the case of QED, whose state-of-the-art is well captured
by exhaustive studies of the fermion-photon vertex
\cite{Bashir:1994az,Bashir:2007qq,Bashir:2011vg,Bashir:2011dp,Kizilersu:2014ela} to implement gauge invariance of physical
observables. In principle, a sound Ansatz for the fermion-photon
vertex should be made in one gauge, say the Landau gauge. The
vertex in any other gauge can then be obtained as the LKFT of the
Landau gauge Ansatz. The sensible implementation of this procedure
guarantees gauge covariance and hence obviates any question about
the gauge dependence of gauge invariant quantities.

We expect the same to be true for QCD, albeit with increasing
complexity. The Landau gauge vertex models in QCD
would transform under some generalized LKFTs to provide an
appropriate model in other linear covariant gauges. An example
which motivates current study takes into account simple 2-point
Green function, more specifically, the Dyson-Schwinger output for
the transversal projection of the gluon propagator for small
values of the gauge parameter. From the available lattice data for
the gluon propagator or its dressing function, it
turns out that there is almost no gauge parameter dependence for
the considered interval of gauge parameter
variation, \cite{Cucchieri:2009kk,Cucchieri:2011aa,Bicudo:2015rma}. This is in sharp contrast
with the Dyson-Schwinger estimates presented in
\cite{Huber:2015ria,Aguilar:2015nqa} which show
sizeable variation with the gauge parameter, compare for example
\cite[FIG.~3 (right panel)]{Bicudo:2015rma} with \cite[Fig.~2
(right panel)]{Huber:2015ria}.

In the Abelian QED case, a frequently adopted strategy is based on
the LKFTs \cite{Landau:1955zz,Fradkin:1955jr} (see also
\cite{Johnson}) which allow us to explicitly transform correlation
functions from one linear covariant gauge with gauge parameter
$\alpha$ to another gauge with parameter $\alpha'$. There is a large body of work which has used these
transformations as a guiding principle toward an improved ansatz
for the three point-vertex and imposing gauge invariant chiral
symmetry breaking, see for example \cite{Burden:1993gy,Bashir:2004hh,Bashir:2005wt,Fernandez-Rangel:2016zac}.
More recently, these transformations have also been studied in the
world line formalism, where LKFTs are generalized to arbitrary
amplitudes in scalar QED, \cite{Ahmadiniaz:2015kfq}.

Similar work in the non-Abelian QCD case has only just
begun~\cite{Aslam:2015nia}, the delay mostly due to the
complexities of the non-Abelian LKFTs. The purpose of the current
article is to write down the formal and natural generalization of
the LKFTs to the non-Abelian case of QCD without jeopardizing
renormalizability.

We study LKFTs by using the gauge invariant fields $A^h$ and
$\psi^h$ as introduced in
\cite{Capri:2015ixa,AminAmin,Capri:2016gut,Capri:2017bfd,Dudal:2017jfw}, see also
\cite{McMullan,Horan:1998im}. We see that these fields, which correspond in
fact to invariant non-local composite gluonic and fermionic
operators, provide us with a rather natural setting to derive both
the known Abelian and the novel non-Abelian LKFTs.

This article is organized as follows: in Section II the
construction of  the gauge invariant fields $A^h$ and $\psi^h$
\cite{Capri:2015ixa,AminAmin,Capri:2017bfd} is briefly summarized,
and then fully exploited in Section III to study the LKFTs for
both gluon and fermion correlation functions. We take a closer
look at the lowest order gluon propagator, retrieving the known
LKFT for the photon propagator. Furthermore, we establish a
relation for the LKFT for the fermion propagator. In Section IV
the LKFTs are derived once more, but now from a different
viewpoint, namely within  the path integral formalism. Fully
exploiting the gauge symmetry of the original classical action,
the path integral allows us to recover the same LKFTs as in
Section III. At last, Section V summarizes our conclusions and
directions for future work.

    \section{ A short summary to the gauge invariant transversal gluon field $A^h_\mu$}
We start from the action \cite{AminAmin,Capri:2017bfd} 
    \begin{equation}\label{action}
        S = S_{FP} + S_f + S_h\,,
    \end{equation}
with the Faddeev-Popov  term given by
    \begin{equation}
        S_{FP} = \int \dd^4 x \left( \frac{1}{4} F_{\mu \nu}^a F_{\mu \nu}^a + \frac{\alpha}{2} b^a b^a + i b^a \partial_\mu A_\mu ^{a} + \bar{c}^a \partial_\mu D_\mu ^{ab} c^b  \right)\,,
     \end{equation}
the matter sector by
    \begin{equation}
        S_{f} = \int \dd^4 x \left(\bar\psi(i\slashed{D}+m_f)\psi\right)\,,
    \end{equation}
and
\begin{equation}
        S_h=\int \dd^4 x \left( \tau^a \partial_\mu A_\mu ^{h,a} + \frac{m^2}{2} A_\mu ^{h,a} A_\mu ^{h,a}+ \bar{\eta}^a \partial_\mu D_\mu^{ab} (A^h) \eta^b\right)\,,
    \end{equation}
where $A^h_\mu$ is defined through
    \begin{equation}\label{stuck}
        A_\mu^h = h^\dagger A_\mu h + \frac{i}{g} h^\dagger \partial_\mu h\,.
    \end{equation}
Here, we set
\begin{equation}\label{h}
h = e^{i g \phi^a T^a}\,,
\end{equation}
with $T^a$ the adjoint generators of SU($N$).  As it is apparent,
the action $S_h$ contains a new  field $\phi^a$, besides the
Lagrange multiplier $\tau^a$ as well as the additional ghost
fields $(\bar\eta^a$, $\eta^a)$\footnote{As underlined in \cite{Capri:2017bfd}, the additional ghosts  $(\bar\eta^a, \eta^a)$ are needed to take into account the Jacobian arising from integration over the Lagrange multiplier $\tau^a$, which gives rise to a delta function of the type $ \delta(\partial_\mu A_\mu ^{h,a})$.}.  All these fields belong to the
adjoint representation.

By construction, the field  $A_\mu^h$ turns out to be transverse,
$\partial_\mu A^h_\mu=0$, and gauge invariant
\cite{Capri:2015ixa,AminAmin}. The transversality of $A_\mu^h$ is
precisely ensured by the presence of the Lagrange multiplier
$\tau^a$. The gauge invariant character of $A^h_\mu$ can be nicely
appreciated from the transformation laws
\begin{equation}\label{htrans}
h\to U^\dagger h\,,\quad\ h^\dagger\to h^\dagger U\,,\quad A_\mu \to A_\mu^U= U^\dagger A_\mu U + \frac{i}{g}U^\dagger \p_\mu U\,,
\end{equation}
with $U$ a generic local SU($N$) transformation.  From eq.~\eqref{stuck}, it follows that
\begin{equation}
(A^h_\mu)^U = A^h_\mu \;.  \label{gah}
\end{equation}

At the quantum level, $A_\mu^h$ is a rather complicated composite operator, nonetheless via the Stueckelberg-like formulation of eq.~\eqref{stuck}, the all order renormalizability of $A_\mu^h$, and thus of its correlation functions, was proven, thanks to the powerful Ward identities underlying the dynamics of the action \eqref{action}, \cite{AminAmin,Capri:2017bfd}. The mass term, $\frac{m^2}{2}A_\mu^{h,a}A_\mu^{h,a}$ , can be put to zero in eq.~\eqref{action}. The  parameter  $m^2$ rather serves to introduce the gauge invariant $d=2$ operator $A_\mu^h A_\mu^h$, well-known from phenomenology \cite{Chetyrkin:1998yr,Gubarev:2000eu,Gubarev:2000nz,Boucaud:2001st,Verschelde:2001ia}.  Though, for our current purposes, we will set $m^2=0$, to restore full equivalence with the standard Yang-Mills QCD action. In that case, the quark-gluon-ghost dynamics of $(S_{FP}+S_f+S_h)$ is equivalent to that $(S_{FP}+S_f)$, as integrating over $\tau, \phi, \bar\eta, \eta$  gives  no more than a unity.  Let us explain  in more detail.  It is important to appreciate the role of the multipliers $\tau^a$ which impose the constraint $\partial_\mu A_\mu ^{h,a}=0$. The latter  can be solved iteratively allowing to express $\phi$ explicitly in terms of $A_\mu$.   More precisely, one finds (see e.g.~\cite{Capri:2016aqq})
  \begin{equation}
\phi =\frac{1}{\partial ^{2}}\partial A+i\frac{g}{\partial ^{2}%
}\left[ \partial A,\frac{\partial A}{\partial ^{2}}\right] +i\frac{g}{%
\partial ^{2}}\left[ A_{\mu },\partial _{\mu }\frac{\partial A}{\partial ^{2}%
}\right] +\frac{i}{2}\frac{g}{\partial ^{2}}\left[ \frac{\partial A}{%
\partial ^{2}},\partial A\right] +\mathcal{O}(A^{3})\,.  \label{phi0}
\end{equation}
In the  expression above, we recognize that the fields $\phi^a$ are the  SU($N$) gauge rotation angles we need to gauge transform a  generic field configuration $A_\mu$ into its transversal, gauge equivalent, configuration $A_\mu^h$. When the $\tau$ is integrated over, i.~e.~we work with the on-shell $\phi$-formulation, the integration over the $\phi$ gives rise to a non-trivial Jacobian, which is lifted into the action via the Grassmann $\bar\eta, \eta$-fields, thereby giving an overall unity.  This will be discussed in more detail in other work.  Notice that this procedure shares great similarity with the introduction of the  unit factor corresponding to the Faddeev-Popov quantization procedure. Also here, both gauge condition and related Jacobian, i.e.~the Faddeev-Popov determinant, are lifted into the action through the introduction of the Lagrange multiplier $b^a$ and of the Faddeev-Popov ghosts $(c^a, {\bar c^a})$.

 Let us also point out that the constraint $\partial_\mu A_\mu ^{h,a}=0$ is what discriminates between the  standard  Stueckelberg action and the formulation \eqref{action}, In particular,  as shown in \cite{AminAmin,Capri:2017bfd},  where a detailed comparison was made with the standard non-renormalizable Stueckelberg theory,  the condition $\partial_\mu A_\mu ^{h,a}=0$ plays a pivotal role in order to ensure the all order renormalizability of the action \eqref{action}.

\section{Derivation of the LKFTs via $A^h$}
In this section, we are going to first re-derive the Abelian LKFTs, followed by the non-Abelian generalization.

A Stueckelberg-based derivation of the LKFTs was already realized
in \cite{Sonoda}, though this analysis is restricted to the
Abelian case, with no clear generalization to the non-Abelian
case.

Up to second order in the coupling  constant $g$, we may write
    \begin{align}
        A_\mu^h &= A_\mu - \partial_\mu \phi + i g \commutator{A_\mu}{\phi} + \frac{ig}{2} \commutator{\phi}{\partial_\mu \phi} - \frac{g^2}{2} \left( A_\mu \phi^2 - 2 \phi A_\mu \phi + \phi^2 A_\mu \right) \nonumber \\
        &+ \frac{g^2}{3!} \left( (\partial_\mu \phi) \phi^2 - 2 \phi (\partial_\mu \phi) \phi + \phi^2 (\partial_\mu \phi) \right) + \mathcal{O}(g^3)\,,
    \end{align}
    or, by denoting the colour index explicitly
    \begin{align}
        A_\mu^{h,a} &= A_\mu^a - \partial_\mu \phi^a - g f^{abc} A_\mu^b \phi^c - \frac{g}{2} f^{abc} \phi^b \partial_\mu \phi^c \nonumber \\
        &- \frac{g^2}{2} D^{abcd} \left( A_\mu^b \phi^c \phi^d - 2 \phi^b A_\mu^c \phi^d + \phi^b \phi^c A_\mu^d \right) \nonumber \\
        &+ \frac{g^2}{3!} D^{abcd}\left( (\partial_\mu \phi^b) \phi^c \phi^d - 2 \phi^b (\partial_\mu \phi^c) \phi^d + \phi^b \phi^c (\partial_\mu \phi^d) \right) + \mathcal{O}(g^3)\,,  \label{exa}
    \end{align}
    with $D^{abcd} = 2 \text{Tr} (T^a T^b T^c T^d)$.

 \subsection{The LKFT for the photon propagator via  $\expval{A^h_\mu(p) A^h_\nu(-p)}$ }
    The  expression \eqref{exa}  can be used to expand   the two-point correlation function $\expval{A^{h,a}_\mu(p) A^{h,b}_\nu(-p)}$. In the Abelian approximation,  i.e.~$A_\mu^h = A_\mu - \partial_\mu \phi $, one immediately obtains
    \begin{align}\label{eq:expvalAhAh}
        \expval{A^{h,a}_\mu(p)A^{h,b}_\nu(-p)}_{\alpha} &= \expval{A_\mu^a(p) A^b_\nu(-p)}_{\alpha} + \expval{A_\mu^a(p) \partial_\nu \phi^b(-p)}_{\alpha} \nonumber \\
        &+ \expval{\partial_\mu \phi^a(p) A_\nu^b(-p)}_{\alpha} + \expval{\partial_\mu \phi^a(p) \partial_\nu \phi^b(-p)}_{\alpha}\,.
    \end{align}
    The two-point correlation functions   $\expval{A_\mu^a(p)\phi^b(-p)}_{\alpha}$ and $\expval{\phi^a(p) \phi^b(-p)}_{\alpha}$ are given by \cite{AminAmin}
    \begin{align}
        \expval{A_\mu^a(p)\phi^b(-p)}_{\alpha} &=  i \alpha \frac{p_\mu}{p^4} \delta^{ab}\,, \\
        \expval{\phi^a(p) \phi^b(-p)}_{\alpha} &=  \frac{\alpha}{p^4} \delta^{ab}\,.        \label{prp}
    \end{align}
    So, eq.~\eqref{eq:expvalAhAh} becomes \footnote{Note that the partial derivations $\partial_\mu$ refer to coordinate space.  A Fourier transformation of the relevant fields has been taken, under the convention $A_\mu(p) = \int A_\mu(x) e^{ip \cdot x} \dd x$, so that $\partial_\mu A_\nu(p)= -i p_\mu A_\nu(p)$.}
    \begin{equation}
        \expval{A^{h,a}_\mu(p)A^{h,b}_\nu(-p)}_{\alpha} = \expval{A_\mu^a(p)A_\nu^b(-p)}_{\alpha} -  \alpha \frac{p_\mu p_\nu}{p^4} \delta^{ab}\,,
    \end{equation}
        or, specifying to the Landau gauge, $\alpha=0$,
    \begin{equation}
        \expval{A^{h,a}_\mu A^{h,b}_\nu}_{\alpha=0} = \expval{A_\mu^a A_\nu^b}_{\alpha=0}\,.
    \end{equation}
     It is worth now to remind that the transverse field $A^h_\mu$ is gauge invariant or, equivalently, BRST invariant, see \cite{AminAmin,Capri:2017bfd}.  From this important feature it follows that the correlation function  $\expval{A^{h,a}_\mu(p) A^{h,b}_\nu(-p)}_{\alpha}$ is BRST invariant as well. As such, it does not depend on the gauge parameter $\alpha$ \cite{AminAmin,Capri:2017bfd}. Therefore,
    \begin{equation}
        \expval{A^{h,a}_\mu A^{h,b}_\nu}_{\alpha} = \expval{A^{h,a}_\mu A^{h,b}_\nu}_{\alpha=0}
    \end{equation}
    and we find
    \begin{equation}
        \expval{A^{a}_\mu(p) A^{b}_\nu(-p)}_{\alpha} = \expval{A^{a}_\mu(p) A^{b}_\nu(-p)}_{\alpha=0} + \alpha \frac{p_\mu p_\nu}{p^4}  \delta^{ab}\,.
    \end{equation}
Said otherwise, we simply recover the LKFT for the photon. Of course, this result can also be easily derived using the underlying BRST invariance of the theory, which ensures that the longitudinal  component of the gluon propagator does not receive any quantum correction, being given by its tree-level approximation.

    \subsection{The LKFT for the gluon propagator via $\expval{A^h_\mu(p) A^h_\nu(-p)}$}
As $A^h_\mu$ is defined also for the non-Abelian case, we can
generalize the foregoing to get  LKFTs for the gluon propagator
via the expansion of
    \begin{equation}\label{ahn}
        \expval{A^{h,a}_\mu(p) A^{h,b}_\nu(-p)}_{\alpha} =\expval{A^{h,a}_\mu(p) A^{h,b}_\nu(-p)}_{\alpha=0}=\expval{A^{a}_\mu(p) A^{b}_\nu(-p)}_{\alpha=0}\,,
    \end{equation}
where, in the last step, we explicitly used that correlation
functions of $A^h_\mu$ reduce to those of $A_\mu$ in the Landau
gauge \cite{Capri:2016gut}.  This property can be appreciated by
realizing that the field $\phi^a$ decouples in the Landau gauge,
as it becomes apparent from the vanishing of the correlation
functions $\expval{A_\mu^a(p)\phi^b(-p)}_{\alpha}$,
$\expval{\phi^a(p) \phi^b(-p)}_{\alpha}$, when $\alpha=0$, see
eqs.~\eqref{prp}.  Notice also that the leading order term of the
expansion of $\expval{A^{h,a}_\mu(p) A^{h,b}_\nu(-p)}_{\alpha}$
will always contain  the gluon propagator in the linear covariant
gauge with gauge parameter $\alpha$.

In what follows,  for the benefit of the reader,  the
next-to-leading order expansion of the l.h.s.~of eq.~\eqref{ahn}
is given, but the contractions of the terms are left open, as this
depends on the precise action one intends to
use\footnote{E.g.~with or without the mass $m^2$ present.}. Also
not included are the necessary vertex insertions in the lowest
order terms  to get the complete $\mathcal{O}(g^2)$ corrections.

Up to second order in the coupling  constant, the expansion of the correlation function  $\expval{A^{h,a}_\mu(p) A^{h,b}_\nu(-p)}$  is found to be\footnote{We suppress the index $\alpha$ from here on.}
        \begin{align}\label{ahn2}
        & \expval{A^{h,a}_\mu(p)A^{h,b}_\nu(-p)}  \\
        &= \expval{A_\mu^a(p) A^b_\nu(-p)} + \expval{A_\mu^a(p) \partial_\nu \phi^b(-p)}
        + \expval{\partial_\mu \phi^a(p) A_\nu^b(-p)} + \expval{\partial_\mu \phi^a(p) \partial_\nu \phi^b(-p)} \nonumber \\
        &+ g f^{bcd} \Big[  - \expval{A_\mu^a (p) A_\nu^c(-p)\phi^d(-p)} + \expval{\partial_\mu \phi^a(p) A_\nu^c(-p)\phi^d(-p)}  \nonumber\\
        & \qquad-  \frac{1}{2} \expval{A_\mu (p)^a \phi^c(-p)\partial_\nu \phi^d(-p)} + \frac{1}{2} \expval{\partial_\mu \phi^a (p) \phi^c(-p)\partial_\nu \phi^d(-p)} \Big] \nonumber\\
        &+ g f^{acd} \Big[ - \expval{A_\mu^c(p) \phi^d(p) A_\nu^b(-p) } + \expval{A_\mu^c(p) \phi^d(p) \partial_\nu \phi^b(-p) } \nonumber\\
        & \qquad - \frac{1}{2} \expval{\phi^c(p) \partial_\mu \phi^d(p) A_\nu^b(-p)} + \frac{1}{2} \expval{\phi^c(p) \partial_\mu \phi^d(p) \partial_\nu \phi^b(-p)} \Big]\nonumber\\
        &+ \frac{g^2}{3!} D^{bcde} \Big[ \expval{A_\mu^a(p) \partial_\nu \phi^c(-p) \phi^d(-p) \phi^e(-p)} -2 \expval{A_\mu^a(p) \phi^c(-p) \partial_\nu \phi^d(-p) \phi^e(-p)} \nonumber\\
        & \qquad + \expval{A_\mu^a(p) \phi^c(-p) \phi^d(-p) \partial_\nu \phi^e(-p)} - \expval{\partial_\mu \phi^a(p) \partial_\nu \phi^c(-p) \phi^d(-p) \phi^e(-p)} \nonumber\\
        & \qquad + 2 \expval{\partial_\mu \phi^a(p) \phi^c(-p) \partial_\nu \phi^d(-p) \phi^e(-p)} - \expval{\partial_\mu \phi^a(p) \phi^c(-p) \phi^d(-p) \partial_\nu \phi^e(-p)} \Big] \nonumber\\
        &+ \frac{g^2}{3!} D^{acde} \Big[ \expval{\partial_\mu \phi^c(p) \phi^d(p) \phi^e(p) A_\nu^b(-p)} -2 \expval{ \phi^c(p) \partial_\mu \phi^d(p) \phi^e(p) A_\nu^b(-p)} \nonumber\\
        & \qquad + \expval{\phi^c(p) \phi^d(p) \partial_\mu \phi^e(p) A_\nu^b(-p)} - \expval{\partial_\mu \phi^c(p) \phi^d(p) \phi^e(p) \partial_\nu \phi^b(-p)} \nonumber\\
        & \qquad + 2 \expval{ \phi^c(p) \partial_\mu \phi^d(p) \phi^e(p) \partial_\nu \phi^b(-p)} - \expval{\phi^c(p) \phi^d(p) \partial_\mu \phi^e(p) \partial_\nu \phi^b(-p)} \Big]\nonumber\\
        &- \frac{g^2}{2} D^{bcde} \Big[ \expval{A_\mu^a(p) A_\nu^c(-p) \phi^d(-p) \phi^e(-p)} - 2 \expval{A_\mu^a(p) \phi^c(-p) A_\nu^d(-p) \phi^e(-p)} \nonumber\\
        & \qquad + \expval{A_\mu^a(p) \phi^c(-p) \phi^d(-p) A_\nu^e(-p)} - \expval{\partial_\mu \phi^a(p) A_\nu^c(-p) \phi^d(-p) \phi^e(-p)} \nonumber\\
        & \qquad + 2 \expval{\partial_\mu \phi^a(p) \phi^c(-p) A_\nu^d(-p) \phi^e(-p)} - \expval{\partial_\mu \phi^a(p) \phi^c(-p) \phi^d(-p) A_\nu^e(-p)} \Big]\nonumber\\
        &- \frac{g^2}{2} D^{acde} \Big[ \expval{A_\mu^c(p) \phi^d(p) \phi^e(p) A_\nu^b(-p) } - 2 \expval{\phi^c(p) A_\mu^d(p) \phi^e(p) A_\nu^b(-p) }\nonumber \\
        & \qquad+ \expval{\phi^c(p) \phi^d(p) A_\mu^e(p) A_\nu^b(-p) } - \expval{A_\mu^c(p) \phi^d(p) \phi^e(p) \partial_\nu \phi^b(-p)} \nonumber\\
        & \qquad + 2 \expval{\phi^c(p) A_\mu^d(p) \phi^e(p) \partial_\nu \phi^b(-p)} - \expval{\phi^c(p) \phi^d(p) A_\mu^e(p) \partial_\nu \phi^b(-p)} \Big]\nonumber \\
        &+ g^2 f^{acd}f^{bef} \Big[\expval{A_\mu^c(p) \phi^d(p) A_\nu^e(-p) \phi^f(-p)} + \frac{1}{2} \expval{A_\mu^c(p) \phi^d(p) \phi^e(-p) \partial_\nu \phi^f(-p)} \nonumber\\
        & \qquad + \frac{1}{2} \expval{\phi^c(p) \partial_\mu \phi^d(p) A_\nu^e(-p) \phi^f(-p)} + \frac{1}{4} \expval{\phi^c(p)\partial_\mu \phi^d(p) \phi^e(-p) \partial_\nu \phi^f(-p)} \Big] + \mathcal{O}(g^3)\,. \nonumber
    \end{align}

 As already remarked, the first term of the expansion, i.e.~$\expval{A_\mu^a(p) A^b_\nu(-p)}$, is nothing but the gluon propagator in linear covariant gauge with gauge parameter $\alpha$.

In work in progress,  expression  \eqref{ahn2}  will be used to
verify the just derived gluonic LKFT in perturbation theory,
thereby extending the results of \cite{Aslam:2015nia}. Comparison can be
made with the known perturbative results in generic linear
covariant gauge \cite{Davydychev:1996pb}.

    \subsection{The gauge invariant fermion fields and associated LKFT}
    In the matter sector, the fermion fields also have a  gauge invariant counterpart, namely \cite{McMullan,Capri:2016aqq} 
    \begin{equation}
        \psi^h = h^\dagger \psi\,,
    \end{equation}
with $h$ being  still defined via eq.~\eqref{h}, using the same
$\phi^a$, but now coupled to the generators of the fundamental
representation. Clearly, $\psi^h$ is  gauge invariant per
construction.  This feature   can be explicitly verified by
combining the gauge transformation of the fermion field
    \begin{equation}
        \psi \to U^\dagger \psi
    \end{equation}
and eq.~\eqref{htrans}.

The renormalizability  of the composite operator $\psi^h$, although not yet fully established, can be achieved along the same lines of the proof of the renormalizability of the operator $A^h$ \cite{AminAmin,Capri:2017bfd}.  Even if being technically challenging, we do not expect major conceptual issues to occur in the fermion sector when compared to the gluon $A^h_\mu$ case.

 In principle, $\psi^h$ can be expanded in  powers of the the $\phi$-field as before,
 yielding
     \begin{equation}
        \psi^h = \psi - i g \phi \psi - \frac{g^2}{2} \phi^2 \psi + \mathcal{O}(g^3)\,.
    \end{equation}

    \subsection{The LKFT for general $n$-point functions}
    Overall,  when the gauge invariance is translated into the corresponding BRST symmetry \cite{AminAmin,Capri:2017bfd}, it turns out that the correlation functions of gauge invariant quantities like, for instance,   $\expval{A^h_\mu \dots \psi^h \dots \bar{\psi}^h \dots}_{\alpha}$, are independent from the gauge parameters.  Therefore,  for a general $n$-point function, it must hold that
    \begin{equation}
        \expval{A^h_\mu \dots \psi^h \dots \bar{\psi}^h \dots}_{\alpha} = \expval{A^h_\mu \dots \psi^h \dots \bar{\psi}^h \dots}_{\alpha'}
    \end{equation}
    as all entering variables are explicitly  gauge-invariant. At first order this becomes
    \begin{align*}
        &\expval{A_\mu - \partial_\mu \phi + i g \commutator{A_\mu }{\phi} + \frac{ig}{2}\commutator{\phi}{\partial_\mu \phi} \dots \psi - ig \phi \psi \dots \bar{\psi} + i g \phi \bar{\psi} \dots}_{\alpha} \\
        &= \expval{A_\mu - \partial_\mu \phi + i g \commutator{A_\mu }{\phi} + \frac{ig}{2}\commutator{\phi}{\partial_\mu \phi} \dots \psi - ig \phi \psi \dots \bar{\psi} + i g \phi \bar{\psi} \dots}_{\alpha'}\,.
    \end{align*}
    Proceeding as before, in the gluon sector, we can always connect to $\alpha'=0$ (i.e.,~the Landau gauge), thereby replacing $A^h_\mu$  by $A_\mu$.

    Let us have a closer look at the fermion sector to illustrate what happens there. We specify to the fermion propagator.  Using the guage symmetry,
    \begin{equation}
        \expval{\bar{\psi}^h(x) \psi^h(y)}_{\alpha} = \expval{\bar{\psi}^h(x) \psi^h(y)}_{\alpha=0}\,,
    \end{equation}
    the transformation of the $\bar{\psi} \psi$-propagator can be expressed as
    \begin{equation}\label{lkfq}
        \expval{\bar{\psi}(x) e^{ig \phi(x)} e^{-ig \phi(y)} \psi(y)}_{\alpha} = \expval{\bar{\psi}(x) e^{ig \phi(x)} e^{-ig \phi(y)} \psi(y)}_{\alpha=0} = \expval{\bar{\psi}(x) \psi(y)}_{\alpha=0}\,,
    \end{equation}
    where we used that $\partial A = 0$ in the Landau gauge.  The relation \eqref{lkfq} can be equivalently written as
    \begin{equation}\label{lkfq2}
        \expval{\bar{\psi}(x)   \psi(y)}_{\alpha}= \expval{\bar{\psi}(x)e^{-ig \phi(x)} e^{ig \phi(y)}\psi(y)}_{\alpha=0}
    \end{equation}
which is nothing else than the conventional LKFT for
the fermion propagator, see for instance
\cite{Landau:1955zz,Fradkin:1955jr,Aslam:2015nia,Johnson}.

    In the standard Abelian works on LKFTs, the r.h.s.~of eq.~\eqref{lkfq2} is usually factorized into
    \begin{equation}\label{lkfq3}
        \expval{\bar{\psi}(x)   \psi(y)}_{\alpha}= \expval{\bar{\psi}(x)\psi(y)}_{\alpha=0}\expval{e^{-ig \phi(x)} e^{ig \phi(y)}}_{\alpha=0}\,,
    \end{equation}
     with
    \begin{equation}\label{lkfq4}
     \expval{\phi(p) \phi(-p)} = - \alpha \frac{1}{p^4}\,.
    \end{equation}
    We will come back in detail to this issue in the next Section.

    \section{LFKTs from the path integral: the Abelian case}
In what follows, we will refresh the direct path integral
derivation of the Abelian LKFT, which is a kind of
rewriting of the original argument provided in
\cite{Landau:1955zz,Fradkin:1955jr} in a more modern language. In
the next section, we will generalize this derivation to the
non-Abelian case, at the cost of adding several complications of
course.

    Consider for now the QED action
    \begin{equation}\label{s1}
        S = \int \dd[4]{x} \left( \frac{1}{4} F_{\mu \nu} F_{\mu \nu} + \bar{\psi} \slashed{D} \psi + i b \partial_\mu A_\mu + \frac{\alpha}{2} b^2 + \bar{c} \partial^2 c + \bar{J_\psi} \psi + \bar{\psi} J_{\bar{\psi}}  \right)\,,
    \end{equation}
    where we included sources for $\psi$ and $\bar{\psi}$ to define the generating functional of Green functions, $Z(J)$, via the path integral\footnote{We will consider here the complete Green functions obtained by differentiating $Z(J)$ with respect to the source $J$. Though, the conclusions immediately go through for the connected Green functions as well when $Z(J)$ is  replaced by the corresponding generator $Z^c(J)$ via the usual identification $Z(J)=e^{-Z^c(J)}$.}
    \begin{equation}
    Z(J)=\int [\d \mu]~e^{-S}\,.
    \end{equation}
    Next, we transform the path integral variables $A$, $\psi$, and $\bar{\psi}$ using the gauge transformation
    \begin{align}
        U &= e^{ie \phi}\,,\\
        A_\mu &\rightarrow A'_\mu = A_\mu - \partial_\mu \phi\,,\\
        \psi & \rightarrow \psi' = U^\dagger \psi
    \end{align}
    and we select
    \begin{equation}
         \phi = -X \frac{1}{\partial^2} \partial_\mu A_{\mu}\,,
    \end{equation}
    where the constant $X$ can still be chosen appropriately, see later. The gluon field transforms as
    \begin{equation}
        A_\mu \rightarrow A'_\mu = A_\mu + X \frac{1}{\partial^2} \partial_{\mu} \partial_\nu A_\nu
    \end{equation}
       and so
    \begin{equation}
        \partial_\mu A'_\mu = (1+X) \partial_\mu A_\mu\,.
    \end{equation}
    When we perform the following transformation  on the Lagrange multiplier $b$:
    \begin{align}
        b & \rightarrow b' = \frac{1}{1+X} b
    \end{align}
    and redefine the gauge parameter via
    \begin{align}
        \alpha &\rightarrow \alpha' = (1+X)^2 \alpha\,,
    \end{align}
     the action, up to its source part, is transformed into itself, except that the gauge parameter $\alpha$  gets replaced by $\alpha'$. Importantly, also the source terms vary, more precisely we end up with 
    \begin{equation}\label{s2}
        S' = \int \dd[4]{x} \left( \frac{1}{4} F_{\mu \nu} F_{\mu \nu} + \bar{\psi}' \slashed{D} \psi' + i b' \partial_\mu A'_\mu + \frac{\alpha'}{2} b'^2 + \bar{c} \partial^2 c + \bar{J_\psi} U \psi' + \bar{\psi}'^\dagger U^\dagger J_{\bar{\psi}} \right)\,.
    \end{equation}
    It is consequently found that the $\phi$-propagator has the expected form \cite{Fradkin:1955jr,Aslam:2015nia,Johnson}
    \begin{align}
        \expval{\phi(p) \phi(-p)}_{\alpha'} &= -\frac{X^2}{(1+X)^2}  \frac{p_\mu}{p^2} \frac{p_\nu}{p^2} \expval{A'_\mu(p) A'_\nu(-p)}_{\alpha'} \\
        &= - \frac{X^2}{(1+X)^2} \frac{p_\mu}{p^2} \frac{p_\nu}{p^2} \alpha' \frac{p_\mu p_\nu}{p^4} \\
        &= - \frac{X^2}{(1+X)^2} \alpha' \frac{1}{p^4} \\
        &= - X^2 \alpha \frac{1}{p^4}\,.
    \end{align}
Starting from any gauge $\alpha$, if we take the limit $X
\rightarrow -1$ we arrive at the Landau gauge $\alpha'=0$, while
the $\phi$-propagator remains proportional  to  $\frac{1}{p^4}$.
This rather singular behaviour is fundamental to correctly
transform the longitudinal projection of the gluon propagator. We
remind here the latter projection is uniquely fixed by means of
the underlying Ward identities,  {\it i.e.} the  BRST invariance
as well as other additional Ward identities defining the class of
linear covariant gauges at the quantum level, see for instance
\cite{AminAmin,Alkofer:2000wg}.

    \subsection{Application to the fermion propagator}
    The relevant partition function is given by
    \begin{equation}
        Z_{\alpha} = \int [\dd{\mu}] e^{-S}
     \end{equation}
    and, after the earlier described path integral variables transformation, by
    \begin{equation}
        Z_{\alpha'} = \int [\dd{\mu}] e^{-S'}\,,
    \end{equation}
    with $S$ and $S'$ given by eq.~\eqref{s1} and eq.~\eqref{s2}. We assume that the measure remains invariant,  a feature which will be proven explicitly in the next Section. Let us  emphasize  that
    \begin{equation}\label{equi}
        Z_{\alpha} \equiv Z_{\alpha'}\,.
    \end{equation}
    The  fermion propagator is found by deriving $Z_\alpha$ with respect to $J_{\bar{\psi}}$ and $\bar{J_\psi}$,
    \begin{align}
        \expval{\bar{\psi}(x) {\psi}(y)}_{\alpha} &= \frac{\delta^2 Z_{\alpha}}{\delta \bar{J_\psi}(y)\delta {J_{\bar{\psi}}}(x)} \\
        &= \int [\dd{\mu}] \bar{\psi}(x) {\psi}(y) e^{-S}\,.
    \end{align}
     Moreover, from  eq.~\eqref{equi}, it is also given by
    \begin{align}\label{equi2}
        \expval{\bar{\psi}(x) {\psi}(y)}_{\alpha} &= \frac{\delta^2 Z_{\alpha'}}{\delta\bar{J_\psi}(y)\delta{J_{\bar{\psi}}}(x)} \\
        &= \int [\dd{\mu}] \bar{\psi}'(x) U^{\dagger}(x) U(y) {\psi}'(y)  e^{-S'}\,.
    \end{align}
    When $\phi$ is a free field, it is evidently possible to factorize
    \begin{equation}
        \expval{\bar{\psi}(x) {\psi}(y)}_{\alpha} = \expval{\bar{\psi}'(x){\psi}'(y)}_{\alpha'} \expval{U^{\dagger}(x)U(y)}_{\alpha'}\,.
    \end{equation}
    We may ask ourselves if this is still the case when the $\phi$-field couples to $A$. The propagator in the new gauge becomes
    \begin{equation}
        \expval{\bar{\psi}'(x) e^{\frac{-i e X}{1+X}\frac{1}{\partial^2}\partial_\mu A'_\mu(x)} e^{\frac{i e X}{1+X}\frac{1}{\partial^2}\partial_\nu A'_\nu(y)}  {\psi}'(y)}_{\alpha'}\,.
    \end{equation}
    In the next step we expand the exponentials. In first order, this becomes the $\expval{\bar{\psi} {\psi}}$-propagator. In second order we obtain, upon inclusion of a single fermion-gauge boson vertex,
    \begin{align}\label{dd1}
        &\int \dd[4]{z} \frac{- i e X}{1+X} (ie\gamma_\nu) \expval{\bar{\psi}'(z) \psi'(y)}_{\alpha'} \expval{\bar{\psi}'(x) {\psi}'(z)}_{\alpha'} \expval{\frac{1}{\partial^2}\partial_\mu A'_\mu(x) A'_\nu(z)}_{\alpha'}
    \end{align}
  which might spoil the  above mentioned factorization. Notice, however, that expression \eqref{dd1} is proportional to
    \begin{equation}\label{dd0}
        \frac{X}{1+X} \alpha' = X \sqrt{\alpha \alpha'}\,.
    \end{equation}
The $\alpha'$ in the l.h.s.~of eq.~\eqref{dd0} arises from the
longitudinal part of the gauge boson propagator,  hidden in the
last factor of eq.~\eqref{dd1}. In the Landau gauge,
i.e.~$\alpha'=0$ from $X\to-1$, this term disappears and we are
effectively able to factorize this expectation value as in
eq.~\eqref{lkfq3}. This will also hold at higher orders, since any
contraction of a  gauge field $A$ from a vertex with a field $A$
lurking in the exponential of $\phi$ will always vanish in the
Landau gauge, similarly to what was just illustrated.

    \subsection{Application to the photon propagator}
    We can also investigate the photon propagator. Therefore, we add the term $\int \dd[4]{x} J_\mu A_\mu$ to the action, so that
    \begin{equation}
        \expval{A_\mu(x)A_\nu(y)}_\alpha = \frac{\delta^2 Z_\alpha}{ \delta J_\nu(y)\delta J_\mu(x)}
    \end{equation}
    in the original gauge.

    This extra source term in the Lagrangian transforms as
    \begin{align}\label{vanafhier}
        J_\mu A_\mu &\to J_\mu (A'_\mu + \partial_\mu \phi)\\
        &= J_\mu \left( A'_\mu - \frac{X}{1+X} \frac{1}{\partial^2}\partial_\mu \partial_\nu A'_\nu \right)\,.
    \end{align}
    From this, we find for the photon propagator
    \begin{align}
        & D^{(\alpha)}_{\mu \nu}(p^2) = \expval{A_\mu(p) A_\nu(-p)}_{\alpha} \\
        &= \expval{\left( A'_\mu(p) - \frac{X}{1+X} \frac{1}{\partial^2} \partial_\mu \partial_\alpha A'_\alpha(p) \right) \left( A'_\nu(-p) - \frac{X}{1+X} \frac{1}{\partial^2} \partial_\nu \partial_\beta A'_\beta(-p) \right)}_{\alpha'} \\
                &= D_{\mu \nu}^{(\alpha')}(p^2) + \left( - 2 \alpha' \frac{X}{1+X} + \alpha' \frac{X^2}{(1+X)^2} \right) \frac{L_{\mu \nu}}{p^2} \\
        &= D_{\mu \nu}^{(\alpha')}(p^2) - \alpha \left( \frac{\alpha'}{\alpha} -1 \right) \frac{p_\mu p_\nu}{p^4}\,.
    \end{align}
We used $\alpha' = \alpha (1+X)^2$ and the standard photon propagator  decomposition in a general linear gauge:
    \begin{align}\label{foton}
        \expval{A_\mu(p)A_\nu(-p)}_\alpha = D_{\mu \nu}^{(\alpha)}(p^2) = \Delta(p^2) P_{\mu \nu} + \frac{\alpha}{p^2} L_{\mu \nu}\,,
    \end{align}
    with the transversal and longitudinal projectors
    \begin{align}
        P_{\mu \nu} &=
        \delta_{\mu \nu} - \frac{p_\mu p_\nu}{p^2}\,, \\
        L_{\mu \nu} &= \frac{p_\mu p_\nu}{p^2}\,.
    \end{align}
Clearly, eq.~\eqref{foton} expresses that only the longitudinal part of the photon propagator is affected by shifting $\alpha \rightarrow \alpha'$, as it is well-known in the Abelian case.

    \section{LFKT from the path integral: the non-Abelian case}
 We now wish to generalize the foregoing path integral derivation of the LKFTs to a non-Abelian gauge theory supplemented with fermion matter.

 We must first establish a general SU($N$) transformation with matrix $U=e^{ig\phi}$   for all fields, namely, gauge, matter and Faddeev-Popov ghosts,  while maintaining the property that $\partial_\mu A'_\mu = (1+X) \partial_\mu A_\mu$. This is  a necessary requirement, as  it will precisely allow for the rescaling of the Lagrange multiplier $b$, and thereby for that of the gauge parameter $\alpha$. As before, $\phi = \phi^a T^a$.

 The   gauge and matter fields transform as
    \begin{align}
        A_\mu &\to A'_\mu = U^{\dagger} A_\mu U + \frac{i}{g} U^{\dagger} \partial_\mu U \label{full}\,,\\
        \psi &\to \psi' = U^\dagger \psi\,,\\
        U &= e^{i g \phi} = 1 + i g \phi - \frac{g^2}{2} \phi^2 + \mathcal{O}({\phi^3})\,.
    \end{align}
    Now, if we let the Faddeev-Popov ghosts transform in the adjoint representation,
    \begin{align}\label{ctrans}
        c &\rightarrow U^\dagger c U\,,\\
        \bar{c} &\rightarrow U^{\dagger} \bar{c} U\,,
    \end{align}
 we obtain
    \begin{equation}
    \bar{c} \partial_\mu D_\mu c \to \bar{c} \partial_\mu D_\mu c + \bar{c} U (\partial_\mu U^\dagger) D_\mu c + \bar c D_\mu c (\p_\mu U) U^\dagger\,.
    \end{equation}
 This variation can be reabsorbed by means of the shift
    \begin{equation}\label{cshift}
    c \rightarrow c' = c  + \frac{1}{\partial_\mu D_\mu} U^\dagger (\partial_\mu U)  D_\mu c + \frac{1}{\p_\mu D_\mu} D_\mu c (\p_\mu U) U^\dagger\,.
    \end{equation}
Doing so, we obtain the original action, but now with the changed gauge parameter $\alpha'$.

Concretely, let us expand the transformation \eqref{full} to second order in the fields,
    \begin{align}
        A'_\mu &= A_\mu - \partial_\mu \phi - i g \phi A_\mu + i g A_\mu \phi + i g \phi \partial_\mu \phi - \frac{i g }{2} \partial_\mu \phi^2 + \mathcal{O}(\text{fields}^3)\\
        &= A_\mu - \partial_\mu \phi + i g \commutator{A_\mu}{\phi} + \frac{i g}{2}\commutator{\phi}{\partial_\mu \phi} + \mathcal{O}(\text{fields}^3)\,,
    \end{align}
    thence we impose that
    \begin{align}
        \partial_\mu A'_\mu &= \partial_\mu A_\mu - \partial^2 \phi + i g \commutator{A_\mu}{\partial_\mu \phi} + ig \commutator{\partial_\mu A_\mu}{\phi} + \frac{ig}{2} \commutator{\phi }{\partial^2 \phi} + \mathcal{O}(\text{fields}^3)\\
        &\equiv (1+X) \partial_\mu A_\mu
    \end{align}
    and so we require
    \begin{equation}
    \partial^2 \phi = - X \partial_\mu A_\mu +  i g \commutator{A_\mu}{\partial_\mu \phi} + ig \commutator{\partial_\mu A_\mu}{\phi} + \frac{ig}{2} \commutator{\phi }{\partial^2 \phi} + \mathcal{O}(\text{fields}^3)\,.
    \end{equation}
    At first order this gives
    \begin{equation}
    \phi = -X\frac{1}{\partial^2} \partial_\mu A_\mu\,,
    \end{equation}
    which is nothing but the Abelian result.    Solving iteratively for $\phi$ in powers of $A_\mu$, we get
    \begin{align}\label{nieuwephi}
        \phi &= -X \frac{1}{\partial^2} \partial_\mu A_\mu - igX \frac{1}{\partial^2} \commutator{A_\mu}{\frac{1}{\partial^2} \partial_\mu \partial_\nu A_\nu} \nonumber \\
        & \qquad - igX \frac{1}{\partial^2} \commutator{\partial_\nu A_\nu}{\frac{1}{\partial^2} \partial_\mu A_\mu} + \frac{igX^2}{2} \frac{1}{\partial^2} \commutator{\frac{1}{\partial^2} \partial_\mu A_\mu}{\partial_\nu A_\nu} + \mathcal{O}(A^3)\,.
    \end{align}
    Using this solution, we can calculate $A'_\mu$ as a function of the original $A_\mu$
    \begin{align}\label{nieuweA}
        A_\mu' &= A_\mu + X \frac{\partial_\mu \partial  A}{\partial^2} - igX \frac{\partial_\mu}{\partial^2}\commutator{\frac{\partial_\nu \partial A }{\partial^2}}{A_\nu} - i g X \frac{\partial_\mu}{\partial^2} \commutator{\frac{\partial A}{\partial^2}}{\partial A} - \frac{i g X^2}{2} \frac{\partial_\mu}{\partial^2} \commutator{\frac{\partial A}{\partial^2}}{\partial A} \nonumber \\
        & \qquad + i g X \commutator{\frac{\partial A}{\partial^2}}{A_\mu} + \frac{igX^2}{2} \commutator{\frac{\partial A}{\partial^2}}{\frac{\partial_\mu \partial A}{\partial^2}} + \mathcal{O}(A^3)\,.
    \end{align}
Note that for the Landau gauge, $X=-1$, this  expression coincides
with the gauge invariant transversal field $A^h_\mu$, see
e.g.~\cite{AminAmin}. In general, $A_\mu'$ will not be transversal
though.

    Denoting the color dependence explicitly, this equation becomes
    \begin{align}\label{nieuweAkleur}
        A_\mu'^a &= A_\mu^a + X \frac{\partial_\mu \partial  A^a}{\partial^2} - igX f^{abc} \frac{\partial_\mu}{\partial^2} \left( \frac{\partial_\nu \partial A^b }{\partial^2} A_\nu^c \right) - i g X f^{abc} \frac{\partial_\mu}{\partial^2} \left( \frac{\partial A^b}{\partial^2} \partial A^c \right) - \frac{i g X^2}{2} f^{abc} \frac{\partial_\mu}{\partial^2} \left( \frac{\partial A^b}{\partial^2} \partial A^c  \right) \nonumber\\
        & \qquad + i g X f^{abc} \frac{\partial A^b}{\partial^2} A_\mu^c + \frac{igX^2}{2} f^{abc} \frac{\partial A^b}{\partial^2} \frac{\partial_\mu \partial A^c}{\partial^2} + \mathcal{O}(A^3)\,.
    \end{align}
    This functional relation can be inverted to find the old fields in terms  of the new.   Using $\partial A' = (1+X) \partial A$ we obtain
    \begin{align}\label{oudeA}
        A_\mu &= A_\mu' - \frac{X}{1+X} \frac{\partial_\mu \partial  A'}{\partial^2} + \frac{igX}{1+X} \frac{\partial_\mu}{\partial^2}\commutator{\frac{\partial_\nu \partial A' }{\partial^2}}{A_\nu} + \frac{i g X}{(1+X)^2} \frac{\partial_\mu}{\partial^2} \commutator{\frac{\partial A'}{\partial^2}}{\partial A'} + \frac{i g X^2}{2(1+X)^2} \frac{\partial_\mu}{\partial^2} \commutator{\frac{\partial A'}{\partial^2}}{\partial A'} \nonumber \\
        & \qquad - \frac{i g X}{1+X} \commutator{\frac{\partial A'}{\partial^2}}{A_\mu} - \frac{igX^2}{2(1+X)^2} \commutator{\frac{\partial A'}{\partial^2}}{\frac{\partial_\mu \partial A'}{\partial^2}} + \mathcal{O}(A^3)\,.
    \end{align}
     Again, in first order we find the Abelian result
    \begin{equation}
    A_ \mu = A'_\mu - \frac{X}{1+X} \frac{\partial_\mu \partial A'}{\partial^2}\,.
    \end{equation}
     Up to second order we find the old $A$-field as a function of the new
    \begin{align}\label{oudeA2}
        A_\mu &= A_\mu' - \frac{X}{1+X} \frac{\partial_\mu \partial  A'}{\partial^2} + \frac{igX}{1+X} \frac{\partial_\mu}{\partial^2}\commutator{\frac{\partial_\nu \partial A' }{\partial^2}}{A'_\nu} - \frac{igX^2}{(1+X)^2} \frac{\partial_\mu}{\partial^2}\commutator{\frac{\partial_\nu \partial A' }{\partial^2}}{\frac{\partial_\nu \partial  A'}{\partial^2}}
        + \frac{i g X}{(1+X)^2} \frac{\partial_\mu}{\partial^2} \commutator{\frac{\partial A'}{\partial^2}}{\partial A'} \nonumber \\
        & \qquad + \frac{i g X^2}{2(1+X)^2} \frac{\partial_\mu}{\partial^2} \commutator{\frac{\partial A'}{\partial^2}}{\partial A'} - \frac{i g X}{1+X} \commutator{\frac{\partial A'}{\partial^2}}{A'_\mu} + \frac{i g X^2}{(1+X)^2} \commutator{\frac{\partial A'}{\partial^2}}{\frac{\partial_\mu \partial  A'}{\partial^2}} - \frac{igX^2}{2(1+X)^2} \commutator{\frac{\partial A'}{\partial^2}}{\frac{\partial_\mu \partial A'}{\partial^2}} + \mathcal{O}(A^3)
    \end{align}
which constitutes the generalization of eq.~\eqref{vanafhier}.
When applied to the source term, we can perform a similar
derivation to get an explicit connection between the gluon
propagator in 2 different linear covariant gauges, parameterized
by $\alpha$ and $\alpha'$.  Since the expression in the r.h.s.~of
eq.~\eqref{oudeA2} is not restricted to terms containing a
space-time derivative  $\p_\mu$ beyond leading order, this implies
that the transformation \eqref{oudeA2} is also  affecting  the
transversal component of the  gluon propagator. In particular,
when transforming to the Landau gauge, it is clear that we will
recover the same transformation law as obtained in Section IIIA.

Given that we have constructed $\phi$ in full generality, we can
also easily construct the non-Abelian transformation law for the
fermion propagator. In fact, the analysis leading to
eq.~\eqref{equi2} can be mostly taken over, thus we find
    \begin{align}\label{equi3}
        \expval{\bar{\psi}(x) {\psi}(y)}_{\alpha} &= \expval{\bar{\psi'}(x) U^{\dagger}(x) U(y) {\psi'}(y)}_{\alpha=0}\\
        &=\expval{\bar{\psi'}(x)  e^{-i g \phi(x)}e^{i g \phi(y)} {\psi'}(y)}_{\alpha=0}\,.
    \end{align}
    As expected, this non-Abelian LKFT law is in perfect agreement with the alternative derivation with the gauge invariant fermion field $\psi^h$ that resulted in eq.~\eqref{lkfq2}. For completeness, in the Landau gauge, the factorisation into
    \begin{align}\label{equi4}
        \expval{\bar{\psi}(x) {\psi}(y)}_{\alpha}  &=\expval{\bar{\psi'}(x)   {\psi'}(y)}_{\alpha=0}\expval{e^{-i g \phi(x)}e^{i g \phi(y)}}_{\alpha=0}
    \end{align}
still holds, following the same logic as in the Abelian case. It
is important to realize here the inherent complication compared to
the Abelian case: the LKF field $\phi$ is now an infinite series,
represented  by eq.~\eqref{nieuwephi}. This is important, in
particular,  for the renormalizability of the whole construction,
see also the comments in Section II and
\cite{AminAmin,Capri:2017bfd}. As such, our construction is more
general than  that  explored in \cite{Aslam:2015nia}, where higher order
corrections to the quark propagator LKFT were explored, though
keeping the Abelian approximation for the fields $\phi^a$. We have
now unraveled that a self-consistent approach requires adding more
and more terms to $\phi^a$ as the perturbative order increases.

Before turning to our conclusions and  giving an outlook to
follow-up work, there is a subtle point we did not address so far.
In order that $Z_\alpha = Z_{\alpha'}$, we used that the action
remains invariant under the applied transformations. At the level
of the path integral, in order for our derivation to be correct,
we also need that the integration measure remains invariant,
        \begin{equation}
    [\dd{\mu}] = [\dd{\mu'}]\,,
    \end{equation}
i.e., that there is no Jacobian. We do not expect  a non-trivial
Jacobian, since we already derived the transformations using the
gauge invariant $h$-fields without encountering any differences
with the re-derivation via path integral tools and deliberate
omittance of the Jacobian.  Though, to be sure, let us also verify
this explicitly.  More precisely, we will show that the
super-Jacobian\footnote{As we have a mix of commuting and
anticommuting variables, we must consider the super-Jacobian
(Berezinian).} of the transformation,
        \begin{equation}
\mathfrak{J} = \left( \frac{A'^a_\mu , c'^a , \bar c'^a}{A^b_\nu, c^b ,\bar c ^b}\right) = \mqty(\mathfrak{A} & \mathfrak B\\ \mathfrak C &  \mathfrak D)\,,
    \end{equation}
is trivial. The superdeterminant is given by
        \begin{equation}
    \text{sdet}~\mqty(\mathfrak{A} & \mathfrak B\\ \mathfrak C &  \mathfrak D) = \det{\mathfrak{A}+\mathfrak{B}\mathfrak{D}^{-1}\mathfrak{C}}\det{\mathfrak{D}}^{-1}\,.
    \end{equation}
The transformation of the gluon fields $A_\mu$ is independent of
$c$ and $\bar c$, resulting in $\mathfrak{C}=0$ and we see that
the superdeterminant collapses to the product of the individual
determinants.

    For the gluons, using eq.~\eqref{nieuweAkleur}, the argument of this determinant can be calculated (in what follows only the colour dependence concerns us)
        \begin{equation}
    \frac{\delta A^{'a}_\mu}{\delta A_\nu ^b} = \delta^{ab} C + f^{abc} D\,,
    \end{equation}
        with ${C}$ a constant and ${D}$ some function of the field $A_\mu$. We can work at the infinitesimal level, thereby using $\det(1+\mathcal{A}) \approx 1 + \Tr \mathcal{A}$. Because of the antisymmetry of $f^{abc}$, this determinant becomes equal to 1. Note that the constant $C$ will drop after normalisation of the expectation value.

The infinitesimal transformation in the ghost sector,
eq.~\eqref{ctrans}, is found to be
        \begin{align}
    \delta c^a &= -i g f^{abc} \phi^b c^c\,,\\
    \delta \bar c^a &= -i g f^{abc} \phi^b \bar c^c \,,
    \end{align}
    and thence the matrix $\mathfrak{D}$ becomes
        \begin{equation}
    \mqty(\delta^{ab} - i g f^{acb} \phi^c & 0 \\ 0 & \delta^{ab} - i g f^{acb} \phi^c)\,,
    \end{equation}
    which also leads to  trivial $1$ when taking the determinant. Finally, the shift of eq.~\eqref{cshift} evidently comes with a trivial Jacobian.


\section{Conclusions and outlook}
We have employed the gauge invariant fields $A^h_\mu$ and $\psi^h$
to provide an alternative way to derive the LKFTs for general
$n$-point correlators. This derivation was first performed for the
Abelian LKFT for the photon and fermion fields. It reproduced the
correct relations as already known from the literature. The
extension to non-Abelian theories was then
presented. To our knowledge, this is the first time in which the
non-Abelian LKFTs have been derived for arbitrary
$n$-point correlators without any approximation.

To lend further credit to the validity of our
non-Abelian LKFTs, we also presented an independent derivation of
the LKFTs, from the viewpoint of the path integral formalism,
leading to exactly the same transformations.

Specifically, considering the gluon and quark propagators for an
SU($N$) non-Abelian gauge theory, such as QCD, leads to the
relations \eqref{ahn}, \eqref{ahn2} and \eqref{equi4}. Although
these non-Abelian LKFTs do look (and are) non-local in nature, we
stress here that our framework can be cast in a
fully local, and even renormalizable formulation. This claim follows from the observation that the gauge
invariant composite operator $A^h_\mu$ is renormalizable, as
discussed in \cite{AminAmin,Capri:2017bfd}. The key
is using the algebraic renormalization formalism based on a
Stueckelberg-like reformulation of $A^h_\mu$, in which case the
field $\phi$ is kept as a basic field with its corresponding
propagator given by eq.~\eqref{lkfq4}. A delicate point is the
potential occurrence of infrared singularities when such
propagator is explicitly used in $d=4$. However,
this can be overcome in a BRST consistent fashion by introducing a
regulating mass in the $\phi$-sector that is to be sent to zero at
the end of any calculation
\cite{AminAmin,Capri:2017bfd,Capri:2016gut}, see also \cite{Capri:2017npq}. Details on this will
be presented in a work in progress, where the one
loop explicit check in terms of Feynman diagrams will be worked
out.

Let us discuss the prospect of applying these non-Abelian LKFTs to
non-perturbative functional studies of QCD, in particular related
to its constituent gluon and quark dynamics, followed by their
role in the Bethe-Salpeter and Faddeev equations,
usually employed to study the bound state spectrum of QCD. In the
short run, a perturbative verification of our formalism is
planned. The gauge invariance of chiral quark condensate,
associated primarily with the quark propagator, may be a next
relatively more involved problem.

In the long run, a comprehensive study of hadronic
observables through DSEs and establishing their strict gauge
invariance would be highly desirable, thus raising this formalism
to a higher level of credibility and acceptance. In this
connection, as already mentioned in the introduction, till now
such efforts have been mostly restricted to the Landau gauge.
Several (constrained) Ans\"atze have been put forward, with
increasing complexity, each time making improved contact with
phenomenology \cite{Cloet:2013gva} and also with underlying QCD
dynamics \cite{Binosi:2014aea}. The validity of such Ans\"atze at
the level of gauge covariance, and ultimately, gauge invariance,
is crucial. An ideal goal is to construct an Ans\"atze in a
generic linear covariant gauge parameterized in terms an arbitrary
value of $\alpha$, either explicitly or through the
defining entities of different Green functions. Not only that such
an Ans\"atze should abide by the key symmetries of QCD but should
also stand firm against any explicit check to what extent
physical quantities are effectively gauge invariant.
This is a prohibitively daunting task, as is evident in the much
simpler QED studies as well see
\cite{Bashir:1994az,Bashir:2007qq,Bashir:2011vg,Bashir:2011dp,Kizilersu:2014ela}.
However, we must realize that even if we now have access to the
non-Abelian LKFTs, our construction has been cast in
a perturbative form, viz.~determined by the infinite series
expansion of the gauge invariant variables $A^h_\mu$ (or $\psi^h$)
in the field $\phi$. An intrinsically non-perturbative setup would
require to work with the matrix field $h$ introduced in
eq.~\eqref{h}, and its exact quantum behaviour. This does not
appear feasible at the moment within our approach. So
we may have to resort to an approximate framework. Given that
e.g.~Dyson-Schwinger equations anyhow require a truncation at
finite order (i.e.~finite number of $n$-point interactions), one
could restrict to the 4-point level expansion, which includes the
3-point and 4-point gluon vertices, the 3-point ghost-gluon vertex
and 3-point quark-gluon vertex. For each of these vertices in
Landau gauge, several results are available in
literature from a variety of sources, see
\cite{Cucchieri:2008qm,Blum:2014gna,Pelaez:2015tba,Cyrol:2016tym,Rojas:2013tza,Binosi:2016wcx,Sternbeck:2017ntv}
for a small and thus incomplete selection. The rigorous formalism developed in the current paper can be applied
to get corresponding vertices in another gauge. At first instance,
this can be done in perturbation theory, thereby extending the
work of \cite{Bermudez:2017bpx,Aslam:2015nia}. This might be more realistic than a priori expected, as it is
conceivable that the relevant non-perturbative infrared physics,
hiding in gauge variant interaction vertices and resulting in
gauge-invariant physical observables, may have the gauge dependent
pieces of a perturbative nature, not necessarily or easily
summable in a closed form. Though seemingly an interesting
viewpoint, it needs closer scrutiny and further
exploration.


This being said, it is well known that the contemporary way to
deal with  issues related to  the gauge covariance is via the
powerful BRST invariance
\cite{Becchi:1974xu,Becchi:1974md,Piguet:1995er,Barnich:2000zw},
or more precisely via its functional representation, the
Slavnov-Taylor identity. From the latter, it is not only possible
to derive various relations between different correlation
functions in a fixed gauge, but also how $n$-point functions vary
in terms of the gauge parameter. The latter relations are encoded
in the Nielsen identities
\cite{Nielsen:1975fs,Breckenridge:1994gs,Piguet:1984js}, which
follow directly from the Slavnov-Taylor identity. Given that the
original LKFTs predate the BRST construction with about 2 decades,
one cannot help but wonder if it would not be possible to
construct such transformations directly from the Nielsen
identities, which after all have the same goal as the LKFTs: a
mathematical way to write down how $n$-point functions change
under a changing gauge parameter.  In recent work
\cite{Capri:2016gut}, the Nielsen identity and its consequences in
relation to the gauge invariant propagator $\expval{A^h_\mu(p)
A^h_\nu(-p)}$ were already discussed. As a corollary, we derived
the Abelian LKFT for the photon from the integrated version of the
photon propagator's Nielsen identity. Moreover, in a
recent work, LKFT have been employed to show the gauge invariance
of the electron pole mass in QED, something that was proved
through the corresponding Nielsen identities earlier
\cite{DrPietro}. We are now aiming at exploring how this can be
generalized to the non-Abelian case, thereby hopefully uncovering
new powerful uses of the Nielsen identities. For example,  when
integrated with respect to the gauge parameter, the Nielsen
identity automatically leads to an exponential factor connecting
the propagators in different linear gauges. Such relation was
hinted at in \cite{Aslam:2015nia} but not yet proven in the
non-Abelian case. This and other matters will be discussed in a forthcoming work. In this context, it is also
interesting to point out that the Nielsen identities were also
explored in \cite{Aguilar:2015nqa} in relation to a dynamical mass
generation in linear covariant gauges in a Dyson-Schwinger
framework, thereby uncloaking certain subtleties that deserve
further attention.

Overall, it should be clear that the current article
is a first preliminary step in our rigorous
formalism that can be extended in several directions, as sketched
above.

\section*{Acknowledgments}
T.~De Meerleer is supported by a KU Leuven FLOF grant.
S.P.~Sorella acknowledges the Conselho Nacional de Desenvolvimento
Cient\'{i}fico e Tecnol\'{o}gico (CNPq-Brazil) and The
Coordena\c{c}\~{a}o de Aperfei\c{c}oamento de Pessoal de N\'{i}vel
Superior (CAPES) for support. A.~Bashir
acknowledges CONACyT and CIC-UMSNH grants CB-2014-242117 and 4.10,
respectively.

\end{document}